\documentstyle[prd,aps,eqsecnum,preprint,tighten,floats]{revtex}
\begin{document}
\draft

\preprint{\vbox{\hfill OHSTPY--HEP--T--96-027}
          \vbox{\vskip0.5in}
          }

\title{The Condensate for $SU(2)$ Yang-Mills Theory in 1+1 Dimensions \\Coupled
to Massless Adjoint Fermions\footnote{to appear in the
proceedings of The Conference on Low Dimensional Field, Theory Telluride
Summer Research Institute, August 1996, Internation Journal of Modern
Physics A}}

\author{Stephen S. Pinsky and Richard Mohr}
\address{Department of Physics, The Ohio State University, Columbus,
OH 43210}

\date{\today}

\maketitle

\begin{abstract}
We consider $SU(2)$ Yang-Mills theory in 1+1 dimensions coupled to massless
adjoint fermions.  With all fields in the adjoint representation the
gauge group is actually $SU(2)/Z_2$, which possesses nontrivial
topology.  In particular, there are two distinct topological sectors
and the physical vacuum state has a structure analogous to a $\theta$
vacuum.  We show how this feature is realized in light-front
quantization, using discretization of $x^-$ as an infrared regulator.  We find
exact expressions for the vacuum states and construct the analog of
the $\theta$ vacuum.  We calculate the bilinear condensate of the model. We argue
that this condensate does not effect the spectrum of the massless theory but gives the
string tenson of the massive theory.
\end{abstract}
\pacs{ }

\section{Introduction}
We shall focus our attention here on $SU(2)$ Yang-Mills theory coupled to
adjoint fermions.  This theory is known to have nontrivial vacuum structure; as
first pointed out by Witten \cite{wit79}.  For gauge group $SU(2)$ the model
has a $Z_2$ topological structure and possesses two distinct vacuum states
which we calculate explicitly. The physical vacuum state is a state analogous to a
$\theta$ vacuum that allows the cluster property to be satisfied.  There is a
nonvanishing bilinear condensate \cite{smi94} which we calculate. We will
argue that the condensate that we find is not related to  the condensate found
in the equal-time theory \cite{les94}. We are interested
in this theory because it is an example of a light-cone quantized theory with a
non-trivial vacuum. This theory is interesting for a variaty of additional
reasons.  It has been shown to have the same massive spectrum as QCD$_{1+1}$ with
two colors of fundamental matter, where the topology is trivial and there is a
unique vacuum state \cite{kus95}.  We will argue
that the condensate does not affect the massive spectrum of the the massless
adjoint theory but is related to the string tension in the theory with massive
adjoint fermions \cite{grk95} 

\section{SU(2) Gauge Theory Coupled to Adjoint Fermions: Basics}
Let us now consider SU(2) gauge theory coupled to adjoint fermions
in one space and one time dimension.  Since all fields transform
according to the adjoint representation, gauge transformations that
differ by an element of the center of the group actually represent the
same transformation and so should be identified.  Thus the gauge group
of the theory is $SU(2)/Z_2$, which has nontrivial topology:
$\Pi_1[SU(2)/Z_2]=Z_2$, so that we expect two topological sectors.
This situation differs from the case when the matter fields are in the
fundamental representation, where the gauge group is $SU(2)$ and the
first homotopy group is trivial.

The Lagrangian for the theory is
\begin{equation}
{\cal L} = - {1 \over 2} Tr (F^{\mu \nu} F_{\mu \nu}) + {i \over 2}
Tr(\bar\psi\gamma ^{\mu} \buildrel \leftrightarrow \over
D_{\mu}\psi)\; ,
\end{equation}
where $D_{\mu} = \partial_{\mu} + ig [ A_\mu,\ \ ] $ and $F_{\mu \nu}
= \partial_{\mu}A_{\nu} - \partial_{\nu} A_{\mu} + ig [A_{\mu},
A_{\nu} ]$.  A convenient representation of the gamma matrices is
$\gamma ^0 = \sigma^2$ and $\gamma^1 = i\sigma^1$, where $\sigma^a$
are the Pauli matrices.  With this choice the Fermi field may be taken
to be hermitian.

The matrix representation of the fields makes use of the $SU(2)$
generators $\tau^a=\sigma^a/2$.  It is convenient to introduce a color
helicity, or Cartan, basis, defined by
\begin{equation}
\tau^\pm\equiv {1\over\sqrt{2}}\left(\tau^1\pm i\tau^2\right) \quad,
\left[\tau^+, \tau^- \right] = \tau^3 \quad,
\left[\tau^3 , \tau^{\pm} \right] = \pm\tau^{\pm}.
\end{equation}
Lower helicity indices are defined by
$\tau_\pm = \tau^\mp$. In terms of this basis, matrix-valued fields are decomposed
as, for example,
\begin{eqnarray}
A^{\mu} &=& A _3 ^{\mu} \tau ^3 + A ^{\mu} _+ \tau^+ + A_-^{\mu}\tau^- \nonumber \\
\Psi _{R/L} &=& \psi _{R/L} \tau ^3 + \phi _{R/L} \tau ^+ +
\phi^\dagger _{R/L} \tau ^-
\end{eqnarray}
where $A^{\mu,\pm}\equiv (A^\mu_1\pm iA^\mu_2)/ \sqrt{2}\;$;
$A^{\mu,\pm} = A^\mu_\mp\;$;
$(A^\mu_+)^\dagger=A^\mu_-$  and  $\phi_{R/L}\equiv
(\Psi^1_{R/L}-i\Psi^2_{R/L})/\sqrt{2}$. 

We shall regulate the theory by requiring that the gauge field
$A^{\mu}$ be periodic and the right-handed Fermi field be antiperiodic
in $x^-$.  The left-handed fermion $\Psi_L$ is taken to be
antiperiodic in the coordinate $x^+$.  In most cases we will not present
explicitly the left handed contributions. These are presented and discussed
in more detail elsewhere \cite{mcp96}.

The Fock space representation for the Fermi fields is obtained by
Fourier expanding $\Psi_R$ on $x^+=0$,
\begin{eqnarray}
\psi _R &=& {1 \over 2 ^{1/4} \sqrt {2L}} \sum_n \left(a_n e^{-ik {^+
_n} x^-} + a{_n ^\dagger} e^{ik{_n ^+}x^-} \right) \\
\phi _R &=& {1 \over 2 ^{1/4} \sqrt {2L}} \sum_n \left(b_n e^{-ik {^+
_n} x^-} + d{_n ^\dagger} e^{ik{_n ^+}x^-} \right)\; .
\end{eqnarray}
and similarly $\Psi_L$ on $x^-=0$ with $a_n, b_n, d_n$ and $x^-$ replaced by
$\alpha_n, \beta_n, \delta_n$ and  $x^+$. Here the sums run over the positive
half-odd integers and
$k{^\pm _n} = n \pi /L$.  The Fourier modes obey the standard anti-commutation
relations
\begin{equation}
\{a{^\dagger _n}, a_m \} = \{ b{_n ^\dagger}, b_m \} = \{ d{_n
^\dagger}, d_m \} = \delta _{n, m}
\label{rhccrs}
\end{equation}
with all mixed anti-commutators vanishing.  
The fermionic Fock space is generated by acting with the various
creation operators to a vacuum state $|0\rangle$.

\section{Current Operators}

The current operators for this theory are
\begin{eqnarray}
J^+ \equiv J^R &=& -{1\over\sqrt 2}[\Psi _{R},\Psi_R]
\label{jplus} \; .
\label{jminus}
\end{eqnarray}
and similarly for $J^- \equiv J^L$.
To avoid confusion, we shall henceforth always write the currents with
$R$ or $L$ in place of the upper Lorentz index and the color helicity
index either up or down with $J_3=J^3$ and $J_- = J^+$.  

These expressions for the currents are
ill-defined as they stand since they contain the product of operators
at the same point. This is a common problem and occurs in the
expression for the Poincar\'e generators as well.  We shall regulate
these expressions by point splitting while maintaining gauge
invariance and then take the splitting to zero after removing the
singularities. The singularities give rise to additional
contributions, so called gauge corrections. We find that the current $J^R$
acquires a gauge correction
\begin{equation}
J^R = \tilde{J^R} - {g \over 2\pi} V\; ,
\label{Ranomaly}
\end{equation}
where $\tilde{J^R}$ is the naive normal-ordered current. The left handed current
has an anomally $-{g \over 2\pi} A$. 

It is convenient to Fourier expand these currents and discuss the the
properties of their components.  We write
\begin{eqnarray}
\tilde{J}^{R,a} &=& \sum_{N=-\infty}^{\infty} C^a_N e^{-i\pi N x^-/L}\;
\end{eqnarray}
where $a$ is a color index and the sums are over the integers. The
coefficients in the expansion for the left handed currents are $D^a_N$. It is
well known that these Fourier components satisfy a Kac-Moody algebra with
level one \cite{goo86}.  We shall discuss this explicitly for the $C_N^a$; an
identical set of relations holds for the
$D_N^a$, with appropriate substitutions.

In terms of the Fock operators, we find,
\begin{eqnarray}
C_N^3 &=& \sum_{n={1 \over 2}}^\infty b^\dagger_n b_{N+n}
-\sum_{n={1 \over 2}}^\infty d^\dagger_n d_{N+n}
-\sum_{n={1 \over 2}}^{N-{1 \over 2}} b_n d_{N-n}  \nonumber  \\
C_N ^+ &=& \sum_{n={1 \over 2}}^\infty a^\dagger_n d_{N+n}
- \sum_{n={1 \over 2}}^\infty b^\dagger_n a_{N+n}
-\sum_{n={1 \over 2}}^{N-{1 \over 2}} d_n a_{N-n}  \nonumber  \\
C_N ^- &=& \sum_{n={1 \over 2}}^\infty d^\dagger_n a_{N+n}
 - \sum_{n={1 \over 2}}^\infty a^\dagger_n b_{N+n}
-\sum_{n={1 \over 2}} ^{N-{1 \over 2}} a_n b_{N-n}\; .
\end{eqnarray}
The negative frequency modes may be obtained by hermitian conjugation: 
$C^3_{-N} = (C^3_N)^\dagger$, $C^+_{-N} = (C^-_N)^\dagger$ and $C^-_{-N} =
(C^+_N)^\dagger$ .

In this Cartan basis, the Kac-Moody algebra takes the form
\begin{eqnarray}
\left [ C^3_N , C^3_M         \right ] &=& N \delta_{N,-M} \\
\left [ C_N^{\pm} , C_M^{\pm} \right ] &=& 0 \\
\left [ C^3_N , C_M^{\pm}     \right ] &=& \pm C_{N+M}^{\pm} \\
\left [ C^+_N , C^-_M         \right ] &=& C^3_{N+M} + N \delta_{N,-M}\; .
\end{eqnarray}
Also notice
that for $N=M=0$ the above algebra is the $SU(2)$ algebra of the
charges.  The algebra satisfied by the $D_N^a$ is of course identical.

\section{Gauge Fixing}
 It is most convenient in light-front field
theory to choose the light-cone gauge $A^+ \equiv V =0$.  This is not
possible with the boundary conditions we have imposed, however it is
permissible to take
$\partial_-V=0$.  In addition, we can make
a further global (i.e., $x^-$-independent) rotation so that the zero
mode of $V$ has only a color 3 component \cite{kap94,pik95},
\begin{equation}
V = v (x^+) \tau ^3 \; .
\end{equation}

At this stage the only remaining gauge freedom involves certain
``large'' gauge transformations, which we shall denote $T^R_N$ and
$T^L_N$ with $N$ any integer:
\begin{eqnarray}
T^R_N &=& \exp\left[ -{ iN \pi \over 2L}  x^-\tau_3\right] \nonumber \\
T^L_N &=& \exp\left[ { iN \pi \over 2L}  x ^+\tau_3\right]\; .
\end{eqnarray}
The combination $T^R_N T^L_N \equiv T_N$ is a gauge freedom of the theory
and an example of the Gribov ambiguity \cite{gri78}.  We can use $T_N$ to bring
$Z_R$ to a FMD, $-1 < Z_R < 0$.  Once this is done all gauge freedom has been
exhausted and the gauge fixing is completed.

The physical degrees of freedom that remain are the Fermi fields
$\Psi_R$ and $\Psi_L$ and the gauge field zero mode $Z_R$, restricted
to the finite interval $-1 <Z_R< 0$.  All other nonvanishing
components of the gauge field will be found to be constrained, as is
usual in light-front field theory.  Because of the finite domain of
$Z_R$, it is convenient to use a Schr\"odinger representation for this
variable.  Thus the states of the theory will be written in the form
\begin{equation}
|\Phi\rangle = \zeta(Z_R)|{\rm Fock}\rangle\; ,
\end{equation}
where $\zeta(Z_R)$ is a Schr\"odinger wavefunction and $|{\rm
Fock}\rangle$ is a Fock state in the fermionic variables.  There
remains the question of what boundary conditions should be satisfied
by the wavefunction $\zeta$.  A careful analysis of the integration
measure forces the wavefunction to vanish at
the boundaries of the fundamental domain $  \zeta(-1)=\zeta(0)=0\; .$

After gauge fixing $T_N$ is no longer a symmetry of the theory, but
there is an important symmetry of the gauge-fixed theory that is
conveniently studied by combining $T_1$ with the so-called Weyl
transformation, denoted by $R$.  Under $R$, $RZ_{R}R^{-1} = -Z_{R}$.
$R$ is also not a symmetry of the gauge-fixed theory, as it takes
$Z_R$ out of the fundamental domain.  However, the
combination $T_1R$ is a symmetry.  We
find
\begin{eqnarray}
T_1R Z_R R^{-1}T_1^{-1} &=& -Z_R-1 \; ,
\end{eqnarray}
so that $T_1R$ maps the FMD $-1<Z_R<0$ onto itself.  In
fact, it represents a reflection of the FMD about its
midpoint $Z_R=-1/2$.

The action of $T^{R/L}_1$ and $R$ on the fermion Fock operators  gives rise
to a spectral flow for the right handed fields,
\begin{eqnarray}
T^R_1 b_n T^{-1R}_1 & = & b_{n-1} \quad , \qquad n > 1/2 \nonumber \\
T^R_1 d_n T^{-1R}_1 & = & d _{n+1} \nonumber \\
T^R_1 b _{1/2} T^{-1R}_1 & = & d {_{1/2} ^\dagger} \nonumber \\
R b_n R^{-1} &=& - d_n  \;, 
\label{trsymm}
\end{eqnarray}
and similarly for the left handed operator.
The $a_n$ and $\alpha_n$ are invariant under both $T^{R/L}_1$ and $R$. From
the behavior of the Fock operator it is straight forward to deduce the behavior
of the elements of the Kac Moody algebra under $T^R$, $T^L$ and R and show that
the algebra is invariant. We shall elaborate on the detailed implications of the
symmetry $T_1R$ when we consider the structure of the vacuum state below.

Finally, let us discuss Gauss' law and determine the rest of the vector
potential in terms of the dynamical degrees of freedom.
Gauss's law in matrix form is in the Cartan basis, this becomes
\begin{eqnarray}
-\partial {^2 _-} A_3 &=& gJ_3^R
\label{gauss3}\\
\nonumber\\
-(\partial _- \pm igv)^2 A_\pm &=& g J_\pm^R\; .
\label{gausspm}
\end{eqnarray}
Note that because of the gauge choice
$J^R$ only acquires a gauge correction to its 3 color component.
All color components of $J^L$
receive a gauge correction.

Eqn. (\ref{gauss3}) can be used to
obtain the normal mode part of $A_3$ on the surface $x^+=0$:
\begin{equation}
A_3 = {gL \over 2\pi^2} \sum_{N \neq 0} { C^3_N \over N^2}
e^{-iN \pi x^-/L}\; .
\end{equation}
The zero mode of Eq.(\ref{gauss3}) requires special attention and is
discussed eleswhere\cite{mcp96}.

Because of the restriction of $Z_R$ to a finite domain and the
boundary condition on $\zeta(Z_R)$, the covariant derivatives appearing
in Eqn. (\ref{gausspm}) have no zero eigenvalues.  Thus they may be
inverted to solve for $A_+$ and $A_-$ on $x^+=0$:
\begin{equation}
A_\pm = {gL \over 2\pi^2} \sum_{N} {C^\mp_N \over \left (N \mp Z_R
\right )^2}  e^{-iN \pi x^-/L}\; .
\end{equation}

\section{Vacuum States of the Theory}
The Fock state containing no particles will be called $\vert
V_0\rangle$.  It is one of a set of states that are related to one
another by $T_1$ transformations, and which will be denoted
$|V_M\rangle$, where $M$ is any integer.  These are defined by
\begin{equation}
|V_M\rangle\equiv (T_1)^M|V_0\rangle\; ,
\end{equation}
where $(T_1)^{-1}=T_{-1}$.  It is straightforward to determine the
particle content of the $|V_M\rangle$.   One finds that
\begin{equation}
\vert V_1 \rangle = d^\dagger_{1/2} \beta^\dagger _{1/2} \vert 0 \rangle
\end{equation}
We will focus here on the Fock states $\vert V_0 \rangle$ and $\vert V_1
\rangle$ since we will find that a  degenerate vacuum will be constructed from
them. A general discussion of $\vert V_M \rangle$ follows similar lines. 

The combination $T_1R$ interchanges the Fock states $\vert V_0 \rangle$ and
$\vert V_1 \rangle$ up to a phase,
\begin{eqnarray}
T_1R \vert V_0 \rangle & = & (phase) \vert V_1 \rangle \\ \nonumber
T_1R \vert V_1 \rangle & = & (phase) \vert V_0 \rangle \; .
\end{eqnarray}
All of our states, as we noted previously, will be constructed from a
Schr\"odinger wave function $\zeta(Z_R)$ and a Fock state. We
construct, in the next section, the Schr\"odinger equation by projecting out the
empty Fock state sector of the Hamiltonian. The energy eigenvalues of the system are
proportional to $2L$, as one would expect, since they correspond to fluctuation of the
flux around the entire spatial volume. We will make a large $L$ approximation and only
retain the ground state wave function $\zeta(Z_R)$
We chose the arbitrary phase such that,
\begin{equation}
T_1R  \zeta(Z_R) \vert V_0 \rangle = e^{i\theta} \zeta(-Z_R-1) \vert V_1 \rangle
\end{equation}
We can now construct the ``$\Theta $ vacuum " for this theory that is invariant
under the $TR$ symmetry,
\begin{equation}
\vert \Omega \rangle = \zeta(Z_R) \vert V_0 \rangle
+ e^{i\theta} \zeta(-Z_R-1) \vert V_1\rangle
\label{omega}
\end{equation}
The state we have 
construct in Eqn. (\ref{omega}) is an eigenstate:
\begin{equation}
T_1R \vert \Omega \rangle= \vert \Omega \rangle .
\end{equation}
It is typically necessary to build the theory on such a vacuum state in order to satisfy
the requirements of cluster decomposition as well.
\section{Energy-Momentum Tensor}
The Poincar\'e generators $P^-$ and $P^+$ have contributions from both
the left handed and right handed fermions with $\Psi_R$ is initialized
on $x^+ = 0$ and propagates in $x^+$ while $\Psi_L$ is initialized on
$x^- = 0$ and propagates in $x^-$. These operators must therefore have
contributions from both parts of the initial-value surface.  Thus
\begin{equation}
P^\pm = \int _{-L} ^L dx^- \Theta^{+\pm} + \int _{-L} ^L dx^+
\Theta^{-\pm}\; ,
\end{equation}
where $\Theta^{\mu\nu}$ is the energy momentum tensor.  

The right handed contribution to $P^-$ takes its most
elegant form by inserting the $J^R$s in term of the $C^N$s. We find
\begin{equation}
P^-_{rh} = {g^2 L \over 4\pi^2} \left[
\sum_{N \neq 0}{C^3_N C^3_{-N} \over N^2} +
\sum_N \left [
{C^+_N C^-_{-N} \over (Z_R+N)^2} + {C^-_{N} C^+_{-N} \over (Z_R-N)^2} \right ]
+\Pi^2_R \right]
\end{equation}
In the above equation $\Pi_R=(2 \pi/g)\partial_+ v$ is the momentum
conjugate to $Z_R$, so that $[Z_R,\Pi_R]=i$.

The object is now to find the lowest-lying eigenstates of the Hamiltonian
$P^-$. The gauge anomaly contribution to $P^-$ can be written as $-{\pi \over
 2L}Q_R^2$. Acting on $\vert V_0 \rangle$ $P^-_{lh}$ gives zero and acting on $\vert
V_1\rangle$ it gives $-\pi/2L$ which exactly cancels the contribution from the
one left handed particle in $\vert V_1 \rangle$. We will therefore omit the
term  $-{\pi \over 2L}Q_R^2$ and $P^-_{lh}$ in the following discussion. It is
convenient to separate $P^-$ into a ``free'' part and an interaction,
\begin{equation}
P^- = {g^2L \over 4 \pi^2} [ \Pi^2_R + P{_0 ^-} + P_I^-]\; .
\label{finalham}
\end{equation}
$P_0^-$ includes all $Z_R$-dependent $c$-numbers and one-body Fock
operators that arise from normal ordering Eq. (\ref{finalham}), and
has the form
\begin{equation}
P{_0 ^-} = C(Z_R) + V (Z_R)\; .
\end{equation}
we find,
\begin{equation}
C(Z_R) = -Z_R\psi\;'(1+Z_R)-\psi(1+Z_R)+Z_R\psi\;'(1-Z_R)-\psi(1-Z_R)-2\gamma
\end{equation}
where $\psi$ is the dervitive of the gamma function and $\gamma$ is the Euler
constant. The
one body operator takes the form
\begin{equation}
V(Z_R) = {\sum_n} ( A_n (Z_R) a{_n ^\dagger} a_n + B_n (Z_R) b{_n
^\dagger} b_n + D_n (Z_R) d{_n ^\dagger} d_n )\; .
\end{equation}
where
\begin{equation}
B_n(Z_R) = \psi\;'(Z_R-n +{1 \over 2})-\psi\;'(-Z_R+n +{1 \over 2}),
\end{equation}
$D_n(Z_R) = B_n(-Z_R)$ and $A_n(Z_R)= B_n(Z_R)+D_n(Z_R)$.
$P_I^-$ is a normal-ordered two body interaction.  We do not display
it here as it is unnecessary for our present purposes.  Note also that
$P_0^-$ itself is invariant under $T_1$ and $R$.
This is not true of $C(Z_R)$ and $V(Z_R)$ individually.

Consider the matrix element of $P^-$ acting a possible vacuum state $\zeta(Z_R)
\vert V_0 \rangle$ and an arbitrary Fock state.  The only non-vanishing matrix
element is
\begin{equation}
\langle V_0 \vert P^- \zeta(Z_R) \vert V_0 \rangle = \epsilon _0 \zeta(Z_R)
\end{equation}
which leads to Schr\"odinger equation for $\zeta (Z_R)$:
\begin{equation}
\Biggl [ - {d^2 \over dZ_R^2} +  C(Z_R) \Biggl ] \zeta(Z_R) = 
\epsilon _0 \zeta(Z_R)\; .
\end{equation}
The ``potential'' $C(Z_R)$  has a minimum at $Z_R
= 0$ and diverges at $Z_R=\pm 1$.  The boundary conditions are $\zeta(0) =0$ and
$\zeta (-1) = 0$. These boundary conditions are the result of a number of studies
\cite{les94,kap94} of the behavior of states at the boundaries of
Gribov regions.  It is straightforward to solve this quantum
mechanics problem numberically. We find that a very good fit to the numberical
solution is
\begin{equation}
\zeta(Z_R) =-5.66 \left (1-Z_R^2 \right)^{1.63} Z_R e^{-0.835 Z_R^2} \; .
\end{equation}
Now let us consider the state $\zeta(-Z_R-1) \vert V_1 \rangle$.
Projecting the matrix element of $P^-\zeta(-Z_R-1) \vert V_1 \rangle$
with $\vert V_1 \rangle $ we find
\begin{equation}
\Biggl [ - {d ^2 \over dZ_R^2} +
\bigl [ C(Z_R) + D_{1/2} (Z_R) \bigr ] \biggr ] \zeta(-Z_R-1) = 
\epsilon \zeta(-Z_R-1)\; .
\end{equation}
{}From the explicit forms of $C(Z_R)$ and $D _{1 / 2} (Z_R)$ it can be shown
that $C(Z_R) + D_{1/2} (Z_R) = C(Z_R+1)$. This is of course just the realization
of the $T_1$ invariance of $P ^-_0$.  This explicitly demonstrates that
$\Psi(-Z_R-1) \vert V_1 \rangle $ is a degenerate ground state of the theory.

\section{The Condensate}

It is generally accepted that QCD in 1+1 dimensions coupled to adjoint
Fermions develops a condensate $\Sigma$. So far $\Sigma$ has only been
calculated in various approximations. It has been calculated in the
large-$N$ limit \cite{koz95} and in the small-volume limit for
$SU(2)$ in Ref. \cite{les94}. Previously we calculated a condensate in the
chiral theory with only right-handed fermions \cite{pir96}. In that calculation it was
the field itself that had a condensate and the result was fundamentally
different from what we are considering here. It is an interesting feature
of these theories that the object that condenses depends on the number of
degrees of freedom in the problem. For $SU(3)$ it is a four fermion operator that
has a condensate. Here we consider the vector theory with both dynamical left- and
right-handed fields and we expect to obtain a condensate for $ {\bar \Psi} \Psi$ .

The two vacuum states $\zeta(Z_R) \vert V_0 \rangle$ and $ \zeta(-Z_R-1) \vert V_1
\rangle$ are both exact ground states. We have a spectral flow associated with the
right-handed and left-handed operators, and thus the two physical
spaces in the fundamental domain  differ by  one
right-handed and one left-handed fermion.  They effectively block
diagonalize $P^-$ sectors. One sector is built on a vacuum with no background
particles and the other built on a state with  background particles
and these two sector only communicate through the condensate. There are no interactions in $P^-$ that
connect these blocks.  Furthermore using the
$T_1R$ symmetry one can show that the blocks are identical. Therefore it is argued
\cite{kus95} that the condensate which is a  matrix connecting these states cannot
affect the massive spectrum of the theory. In ref. \cite{grk95} it is pointed out
that if we allow a small mass $m$ for the adjoint fermions the theory becomes
confining with a string tension $\sigma$ that is related to the condensate,
\begin{equation}
\sigma =2 m \Sigma.
\end{equation}

To calculate the condensate $\Sigma$ let us consider $ Tr(\bar {\Psi}(0) \Psi(0))$ and
retain only the contribution of $d_{1/2}$ and $\beta_{1/2}$ particles which
could give a non-zero contribution,
\begin{equation}
Tr(\bar {\Psi}(0) \Psi(0)) = - {i \over 2L \sqrt{2}} 
\left [ d_{1/2}^\dagger \beta_{1/2}^\dagger +d_{1/2}\beta_{1/2} \right]... \;.
\end{equation}
Taking matrix elements with $\vert \Omega \rangle$ we find that only the cross
terms contribute. We find for the vacuum expectation value of 
${\bar \Psi} \Psi$ from Eqn. (\ref{omega}):
\begin{equation}
\Sigma  =
\langle \Omega \vert  Tr( {\bar \Psi(0)} \Psi(0)) \vert \Omega \rangle 
= -{sin(\theta) \over
2L \sqrt{2} } \int_{-1}^0 \zeta(Z_R)\zeta(-Z_R-1) dZ_R.
\label{condensate}
\end{equation}
Evaluating the integral numerically we find,
\begin{equation}
\Sigma ={sin(\theta) \over 2L} .656
\end{equation}
We see that this expression behave like $1/L$.
This is a common result for discrete light cone calculations and is found even in
the Schwinger model where the exact result is known not to have this behavior. This
a result of the crude treatment of the small $p^+$ region that occurs in this
method. The region $p^+=0$ is generally very singular and it is believed that in
the limit
$L \rightarrow \infty$ one encounters these singularities and they cancel the
vanishing $1/L$ leaving a finite result. We also note that the only other
calculation
\cite{les94} of this quantity, done in different gauge, in a small volume
approximation and discretized and quantized at equal time, finds this $1/L$
behavior.
 
Kutasov and Schwimmer \cite{kus95} have pointed out
that there are classes of theories that have the same massive spectrum.  For
example $QCD_{1+1}$ coupled to two flavors of fundamental fermions should have the
same massive spectrum as the theory we have considered here. However the theory with
fundamental fermions will not have a condensate. The pimary conditions
for this universality is the decoupling of the left and right handed fields.  

\section{ Conclusions }

We have shown that in QCD coupled to chiral adjoint fermions in two dimensions the
light-front vacuum is two-fold degenerate as one would expect on general grounds. 
The source of this degeneracy is quite simple.  Because of the existence of Gribov
copies, the one gauge degree of freedom, the zero mode of $A^+$, must be restricted
to a FMD. The domain of this variable, which after normalization we call $Z_R$, is
bounded by the integers.  Furthermore there is a
$T_1R$ symmetry which is effectively a reflection about the midpoint of the FMD. The
$T_1R$ symmetry operator acting on the Fock vacuum generates a second degeneracy
vacuum. Since $T_1$ generates a spectral flow for the fermions this second vacuum
states contains one left-handed and one right-handed fermion each with momentum
${\pi \over 2L}$, but the state can be shown to still have
$P^+=P^-=0$.

We form the analog of a $\theta$ vacuum from these two-fold degenerate vacuum states
which respects all of the symmetries of the theory.  We find that $Tr({\bar
\Psi} \Psi)$ has a vacuum expectation value with respect to this
$\theta$ vacuum and we find a exact expression for this condensate $\Sigma$. It is
unlikely that the condensate we find here is equivalent to the theories that have
been studied in the equal-time formulation \cite{koz95,les94}, because the infrared
regulator used there appears to couples the left and right-handed fermions.

Since $P^-$ is block diagonal in our degenerate vacuum states the condensate does
not affect the massive spectrum of the theory and the theory has no massless bound
state. However this condensate is proportional to the string tension
\cite{grk95} where the adjoint Fermions are given a small mass.

\acknowledgments
\noindent
The work  was supported in part by a grant from the US Department of Energy.
S.S.P. would like to thank Dave Robertson and Gary McCartor for many helpful
conversations.

\end{document}